\documentstyle[prb,aps,epsf,multicol]{revtex}

\begin{document}

\draft
\title{ Particle on a polygon: Quantum Mechanics }
\author{Rajat Kumar Pradhan\cite{rajat}}
\address{Vikram Deb College, Jeypore 764 001, Orissa, India}
\author{Sandeep K. Joshi\cite{jos}}
\address{Institute of Physics, Sachivalaya Marg, Bhubaneswar 751 005,
India}

\maketitle

\begin{abstract}
  
  We study the quantization of a model proposed by Newton to explain
  centripetal force namely, that of a particle moving on a regular
  polygon. The exact eigenvalues and eigenfunctions are obtained. 
   The quantum
  mechanics of a particle moving on a circle and in an infinite
  potential well are derived as limiting cases.
 
\end{abstract}
\begin{multicols}{2}
\section{Introduction}

The model of a particle bouncing off a circular constraint on a
polygonal path was originally devised by Newton to motivate the
concept of centripetal force\cite{chandra} associated with circular
motion of the particle. It serves the precise purpose of highlighting
the notion of a force continuously operating on the particle on a
circle as the limiting case of motion on an inscribed polygon with the
corners acting as force centers. However, the intuitive simplicity of
the circle case has contributed to the consistent ignoring of Newton's
original model.  Recently the problem has received some attention
\cite{ancin,herman} mostly from the historical perspective. No
quantum mechanical treatment of the problem is available in the
literature.  The quantum mechanics of the polygon model will be of
much pedagogical value for introductory courses in quantum theory
inasmuch as it is an illustration of the fact that the symmetry (the
N-fold discrete rotational symmetry in the present case) of a problem
is fully reflected in the wavefunction.

In this work we first reduce the Lagrangian for the system by using
the constraint equation and then use a suitable generalized coordinate
to bring the Hamiltonian to the free particle form. The resulting
eigenvalues are shown to reduce to those of a particle on a circle in
the limit as the number of sides $N \rightarrow \infty$ and to those
of a particle in an infinite potential well in the $N=2$ case. 


\section{The Classical Hamiltonian}

\begin{figure}
\protect\centerline{\epsfxsize=8.5cm \epsfbox{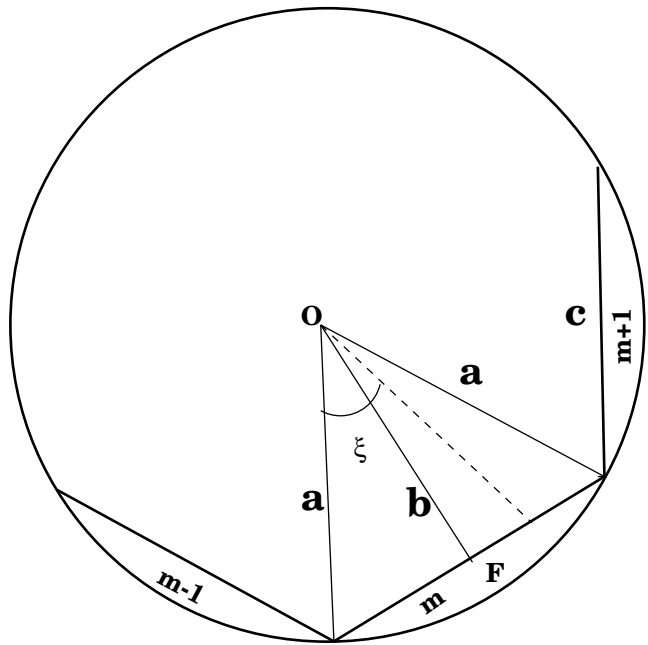}}
\caption{The regular polygon circumscribed by a circle of radius $a$.}
\label{fig1}
\end{figure}

We begin by constructing the full Lagrangian for a particle of unit
mass constrained to move on an $N$-sided regular polygon. The $N$-gon
can be parametrized by

\begin{equation}
\label{rcoord}
r=b~sec(\xi-(2m-1)\pi/N),~~~~~\frac{2(m-1)\pi}{N} \leq \xi \leq
\frac{2m\pi}{N}
\end{equation}

\noindent where $m=1,...N$ labels the sides of the polygon. As shown in
Fig. \ref{fig1}, $b=a~cos(\pi/N)$ is the length of the normal from the
center of the circumcircle (of radius $a$) to the side. The length of
a side is then given by $c=2a~sin(\pi/N)$.

The Lagrangian will be given by

\begin{equation}
\label{lagran}
L=\frac{1}{2}{\dot r}^2 + \frac{1}{2}r^2{\dot \xi}^2+\lambda
(r-b~sec(\xi-(2m-1)\pi/N))
\end{equation}

\noindent where $\lambda$ is a Lagrange multiplier implementing the
constraint

\begin{equation}
\label{const}
\Omega=r-b~sec(\xi-(2m-1)\pi/N)=0
\end{equation}

To reduce the Lagrangian to the only relevant degree of freedom i.e.,
$\xi$, we employ the constraint directly to bring it to the form

\begin{equation}
\label{reduce}
L_r=\frac{1}{2}b^2sec^4(\xi-(2m-1)\pi/N){\dot \xi}^2
\end{equation}

The momentum conjugate to the coordinate $\xi$ is 

\begin{equation}
\label{pxi}
P_{\xi}=\partial
L_r/\partial {\dot \xi} = b^2sec^4(\xi-(2m-1)\pi/N){\dot \xi}.
\end{equation} 

Finally, the classical reduced Hamiltonian is given by

\begin{equation}
\label{clHamilt}
H(\xi)=\frac{P_{\xi}^2}{2b^2}cos^4(\xi-(2m-1)\pi/N)
\end{equation}

\section{Quantization}

A closer look at the Hamiltonian Eq. \ref{clHamilt} reveals that if we
introduce a new generalized coordinate

\begin{equation}
\label{qcoord}
q=b~tan(\xi-(2m-1)\pi/N)
\end{equation}

\noindent it reduces to the simple free particle form

\begin{equation}
\label{Hamilt}
H(q)=\frac{1}{2} {\dot q}^2
\end{equation}

\noindent Following the usual Schr\"odinger prescription for quantization, the
quantum canonical momentum is $P_q=-i\hbar\partial/\partial q=\hbar
k$. The Hamiltonian is therefore the usual free particle Schrodinger
operator

\begin{equation}
\label{schop}
{\cal H}= -\frac{\hbar^2}{2}\frac{\partial^2}{\partial q^2}.
\end{equation} 

A more formal approach to the quantization of such a Hamiltonian Eq.
\ref{clHamilt} would be to construct the Laplace-Beltrami operator in
terms of the properly constructed quantum momentum operator
\cite{rabin}. We emphasize that the same quantum Hamiltonian is
obtained by this procedure also.

The free particle solutions are now given by the plane waves

\begin{equation}
\label{solut}
\psi(q)=A~e^{\pm ikq}
\end{equation}

\noindent Reverting back to the $\xi$ variable we have 

\begin{equation}
\psi_m(\xi)=A~e^{ikb~tan(\xi - (2m-1)\pi/N)}
\end{equation}

\noindent for the wavefunction in the $m^{th}$ side. The normalization is
obtained from

\begin{equation}
\label{norm}
\sum_{m=1}^{N} \int_{2(m-1)\pi/N}^{2m\pi/N} |\psi_m(\xi)|^2 d\xi~=~1
\end{equation}

whence $A=1/\sqrt{2\pi}$.

The boundary condition following from the singlevaluedness of the
wavefunction

\begin{equation}
\label{bcond}
\psi_1(\xi=0)~=~\psi_N(\xi=2\pi)
\end{equation}

\noindent leads to the quantization condition

\begin{eqnarray}
\label{qcond}
kb~tan(\pi/N)&=&n\pi \nonumber \\
& or & \nonumber \\
ka~sin(\pi/N)&=&n\pi
\end{eqnarray}

\noindent The energy eigenvalues are now given by 

\begin{equation}
\label{energy}
E_n~=~\frac{n^2\pi^2\hbar^2}{2a^2sin^2(\pi/N)}.
\end{equation}

\noindent We note that the above quantization conditions are also obtained
by considering symmetric and antisymmetric solutions about the
midpoint of a side, $\xi=(2m-1)\pi/N$ namely,

$$\psi_{m}^s(\xi)~=~A_s~cos(kb~tan(\xi-(2m-1)\pi/N))$$

and

$$\psi_{m}^a(\xi)~=~A_a~sin(kb~tan(\xi-(2m-1)\pi/N)) $$

\noindent and imposing the periodic boundary conditions

$$\psi_{m}^{s,a}(\xi)~=~\psi_{m+1}^{s,a}(\xi+2\pi/N).$$

\begin{figure}
\protect\centerline{\epsfxsize=8.5cm \epsfbox{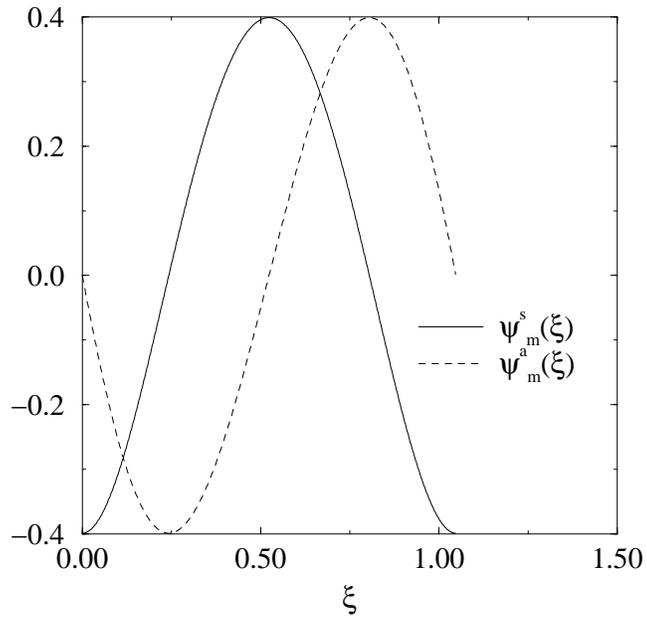}}
\caption{Symmetric ($\psi^s_m(\xi)$) and antisymmetric ($\psi^a_m(\xi)$)
ground state ($n=1$) wavefunctions for the case of hexagon ($N=6$).}
\label{fig2}
\end{figure}

In Fig. \ref{fig2} we show the solutions for a hexagon ($N=6$).

\section{The Circle ($N \rightarrow \infty$) Limit}

The familiar example of a particle moving on a circle\cite{atkins}
would correspond
to the limit $N \rightarrow \infty$ for the polygon model. In this
limit, as each side of the polygon reduces to just a point, it is
necessary to redefine the conjugate variables as $\phi = N\xi$ for the
angle variable and $l=ka/N$ for the angular momentum in units of
$\hbar$. We see that the eigenvalues now become $E_n=n^2\hbar^2/2a^2$
and the eigenfunctions reduce to the well known solutions
$\psi(\phi)~=~A~e^{\pm i n \phi}$ for a free particle on a circle.

%

\section{The Infinite Potential Well ($N=2$) Case}

In the interesting case of $N=2$, classically the polygon reduces to
just a segment of length $2a$ traversed in both directions as
discussed very lucidly by Anicin \cite{ancin}. Quantum mechanically
this is equivalent to a particle confined in an infinite potential
well of width $2a$. Indeed, the eigenvalue equation Eq. \ref{energy}
reduces to the familiar eigenvalues

\begin{equation}
\label{infwell}
E_n=\frac{n^2\pi^2\hbar^2}{8a^2}.
\end{equation}

\noindent Note that in the above we have effected the replacement of
wavevector $k$ by $k/2$ to take care of the two-fold reflection
symmetry which is inherent in the reduction of the polygon to the
segment of length $2a$.  We remark that since the parameterization Eq.
\ref{rcoord} of the polygon does not survive down to the $N=2$ case
for obvious reasons, the eigenfunctions cannot be expressed in terms
of the variable $q$ or $\xi$. The eigenvalue equation Eq.
\ref{energy}, nevertheless, is robust and gives the true eigenvalues.

\section{Summary}

We have quantized Newton's polygon model and derived the general
eigenvalues and eigenfunctions. We have recovered the quantum mechanics in
the circle limit of the polygon. Moreover, the energy eigenvalues for
the $N=2$ case have been identified with those of a particle in an
infinite potential well. In the light of the above results it would be
interesting exercise for students of introductory courses on quantum
mechanics to look into the polygon analogues of the well studied
problems of Stark effect for rotor\cite{pauling}, Aharonov-Bohm effect
for a particle on a circle etc..

\acknowledgments 

We gratefully acknowledge fruitful discussions with V. V. Shenoi, and J.
Kamila. One of the authors (RKP) would like to thank the Institute of
Physics, Bhubaneswar, for hospitality where a part of the work was carried
out.

\end{multicols}

\begin{figure}
\protect\centerline{\epsfxsize=8.5cm \epsfbox{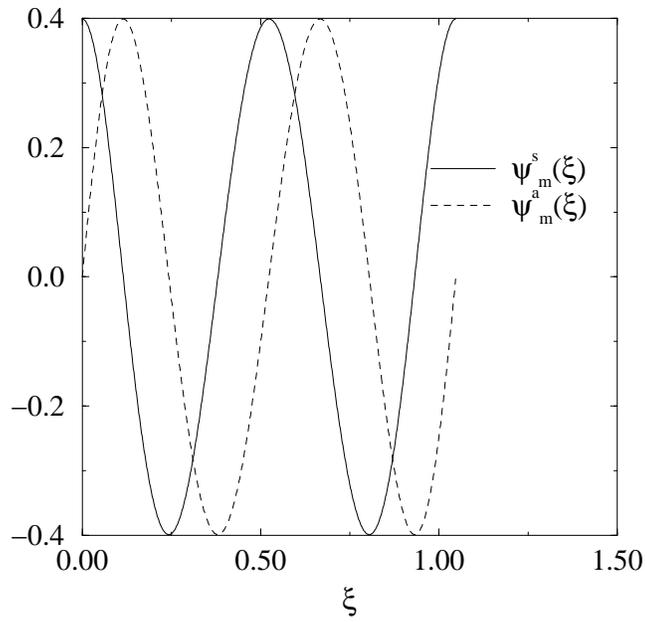}}
\caption{Symmetric ($\psi^s_m(\xi)$) and antisymmetric ($\psi^a_m(\xi)$)
first excited state ($n=2$) wavefunctions for the case of hexagon ($N=6$).}
\label{fig3}
\end{figure}
\end{document}